\documentclass[12pt,a4paper]{article}
\title{Centrality in Production Networks and International Technology Diffusion\footnote{\;I thank Naoto Jinji, Keiko Ito, Sawako Maruyama, and participants of the Japanese Economic Association, the Japan Society of International Economics, and the Kansai Branch workshop of the Japan Society of International Economics for their helpful comments and suggestions on earlier versions of the paper.}}
\date{\today}
\author{\;Rinki Ito\thanks{Graduate School of Economics, Kyoto University, Yoshida Honmachi, Sakyo-ku, Kyoto (Email: itou.rinki.73r@st.kyoto-u.ac.jp).}}

\usepackage{amsmath,amssymb}
\usepackage[dvipdfmx]{graphicx}
\usepackage[dvipdfmx]{color}
\usepackage{dcolumn}
\usepackage{rotating}
\usepackage{caption}
\usepackage{pdflscape}
\usepackage{comment}
\usepackage{setspace}
\usepackage{geometry} 
\usepackage{booktabs}
\usepackage{lscape}
\usepackage[normalem]{ulem}
\usepackage{titlesec}
\usepackage{tabularx}
\usepackage{here}
\usepackage[subrefformat=parens]{subcaption}
\titleformat*{\section}{\large\bfseries}
\titleformat*{\subsection}{\normalsize\bfseries}
\useunder{\uline}{\ul}{}
\begin{document}
\maketitle

\begin{abstract}
This study examines whether the structure of global value chains (GVCs) affects international spillovers of research and development (R\&D). Although the presence of “hub” countries in GVCs has been confirmed by previous studies, the role of these hub countries in the diffusion of the technology has not been analyzed. Using a sample of 21 countries and 14 manufacturing industries during the period 1995-2007, I explore the role of hubs as the mediator of knowledge by classifying countries and industries based on a “centrality” measure. I find that R\&D spillovers from exporters with High centrality are the largest, suggesting that hub countries play an important role in both gathering and diffusing knowledge. I also find that countries with Middle centrality are getting important in the diffusion of knowledge. Finally, positive spillover effects from own are observed only in the G5 countries.
\vspace{5mm}\\
\textbf{JEL-Classification}: F14, L60, O33, O47
\end{abstract}
\hspace{3mm}
 \clearpage
 \section{Introduction}\label{sec1}
\hspace*{3mm}Where do the R\&D spillover effects come from? Starting with Coe and Helpman (1995), the study of international R\&D spillovers has developed over the last three decades, both theoretically and empirically, and have shown that trade is an important channel in the propagation of knowledge. Although several previous studies have confirmed that countries adopt foreign technologies through trade, and this affects productivity, it only investigates the overall effects from abroad. The export is, at best, classified as G7 countries. After all, previous studies do not answer Keller's (1998) question about the possibility that the country from which technology is imported is inconsequential. To answer his question, we need to dig deeper into the connections between the countries.\\
\hspace*{3mm}Trade has increased over the years. However, it has not increased evenly in each country, and has been concentrated in a few countries alone (Criscuolo and Timmis, 2018a).
Additionally, a structure exists whereby many countries trade with the hub in their respective regions, and the hubs trades with the hubs in other regions. Therefore, it is obvious that there would exist a difference in the spillover effects produced between the hubs and the non-hubs. This is because the hub country is responsible for transmitting the technology of other countries, and the indirect effects are likely to be larger than those in the non-hub countries. Although no study has identified and examined the centrality of the export side, I consider that the hubs, which distort the network structure in trade, are important for the diffusion of knowledge.\\
\hspace*{3mm}In fact, recent studies in macroeconomics have found that the manner of interaction of a firm or industry with others in the production network is important for the transmission of shocks. In particular, the presence of large hubs in production networks facilitates the propagation of shocks that are considered localized disturbances (e.g., Gabaix, 2011; Acemoglu et al., 2012; Carvalho, 2014; Acemoglu et al., 2016; Boehm et al., 2019; Carvalho et al., 2021). This is because the hub firm trades with several firms. However, conversely, it means that transaction with such firms indirectly benefits from the technology of other firms.\\
\hspace*{3mm}Motivated by these episodes, I ask the following research questions: How does the structure of the GVCs affect the international spillover effect of knowledge? In particular, are the international R\&D spillover effects from trading with high centrality countries greater than trading with peripheral countries? Are the central countries important in technology diffusion? In addition, are there any differences in the spillover effects due to the differences in properties on the import side? To explore the role of the central country in the diffusion of technology, I use the World Input-Output Database (WIOD) that is the global input-output table, and identifies countries that are central in both the backward and forward linkages (i.e., key buyer and key supplier), following Criscuolo and Timmis (2018a).\footnote{\;Centrality computed here is not an absolute measure, but rather a relative measure; relative to other country-industries in the network. An advantage of this is that the centrality metric has a consistent unit of measurement over time. However, as Criscuolo and Timmis (2018a) say, it does require care to interpret changes in centrality over time, meaning increasing/decreasing influence relative to others in the network. For example, increasing centrality could, in principle, be because you are more influential in absolute terms or because peers in the network have become less influential.} Since the key buyer can be considered important in gathering knowledge from other countries, and the key supplier can be considered important in diffusing knowledge to other countries, I consider that a country that has both the characteristics is important for diffusing other countries' technology. Therefore, any country benefits more by trading with a high centrality country rather than a low centrality country. Lumenga-Neso et al. (2005) show that indirect effects are as important as direct effects in international spillovers. Thus, I consider that the trade with hub countries, which generate more indirect demands due to trade with several countries, will have large spillover effects than peripheral countries. Additionally, I consider that the importers are more likely to benefit from the spillover effect if they are in an environment that facilitates trade with a hub country, while countries with high technological capabilities are less susceptible to spillover effects from abroad. However, since the hub country of this study is central in terms of both ``imports'' and ``exports, '' it is considered to have certain technological capabilities.\footnote{\;This is probably because high productivity is required for exports (Bernard and Jensen, 1999; Clerides et al., 1998; Melitz, 2003).} Therefore, for the latter, trade with high centrality may generate spillover effects.\\
\hspace*{3mm}In my empirical analysis, I examine the hypothesis using a sample of 21 countries and 14 manufacturing industries from 1995 to 2007. Following Nishioka and Ripoll (2012), I decompose the R\&D contents into domestic and foreign. I also analyze the relationship between international R\&D spillovers and the structure of the GVCs by splitting my sample as follows.
First, I decompose the importer into two categories based on geographical and technological aspects (All samples, European countries, and North American countries; G5\footnote{France, Germany, Japan, the U.K, and the U.S.} and non-G5 countries).
Second, to explore the difference in the effects between the exporter's centrality, I divide the countries into three categories: high, middle, and low. The top 33\% countries are high centrality, the next top 33\% countries are middle centrality, and the rest are low centrality in each sector in each year, respectively.
Finally, I focus on the difference of centrality in each industry, and further split exporters. Centrality is computed as a relative value across all trades, thus, even if a country belongs to the high centrality group, it could not be considered a hub in overall trade. If the trade volume of the industry is small, its centrality will naturally be small, and its importance in knowledge diffusion will be reduced. Therefore, to distinguish, I classify the 14 industries into two categories: ``Group A'' and ``Group B (more centralized industry)''.\\
\hspace*{3mm}The main findings of my empirical analysis are as follows: First, the impacts of foreign R\&D contents from high centrality exporters are the highest, and positively significant. These exporters are important as mediators in gathering and diffusing knowledge. Second, trade with middle centrality exporters also have significantly positive effects on domestic total factor productivity (TFP). In particular, the increase in direct and indirect demand from middle centrality countries since 2002 is believed to have contributed to the results. Finally, it seems that only technologically-advanced countries (G5) are affected by the spillover effect from own. Additionally, looking at the share of domestic and foreign R\&D contents, the share of domestic content is much higher in the G5 countries than in the other countries. This result suggests that the proportion of the R\&D contents flowing into a country is important for R\&D spillovers.\\
\hspace*{3mm}This study proceeds as follows. In Section 2, I discuss the conventional research and issues of the impact of international R\&D spillovers. In Section 3, I explain the data, variables, and classifications that I use in the empirical analysis. Section 4 illustrates the empirical strategy. The results of the estimations and the robustness check are reported in Section 5. Section 6 concludes the study.

\section{Literature Review}\label{sec2}

\hspace*{3mm}There has been a long history of study on the impact of foreign R\&D on productivity through trade. It has been studied from both theoretical and empirical perspectives, especially in the empirical study, pioneered by Coe and Helpman (1995) (henceforth, CH). They find that a country's TFP depends on its domestic R\&D stock and on the R\&D stock of the foreign countries. They decipher this by using the import-weighted sums of the trade partners' cumulative R\&D expenditure. The framework for several analysis of subsequent R\&D spillover studies at country or industry levels has been estimated according to CH's model (e.g., Xu and Wang, 1999; Coe et al., 2009).\footnote{\;Keller (2002a, b) use Non-linear Least Squares (NLS). He examines the geographical effect on spillovers, and the link between productivity, and domestic and foreign R\&D at the industry level, respectively. Fracasso and Vittucci Marzetti (2015) modify Keller's NLS model to apply threshold regression. To answer Keller's (1998) question, they study whether the spillovers are relatively stronger (though not necessarily in a proportional way) where trade relations are relatively more intense}\footnote{\;Acharya and Keller (2009) use value-added as independent variable, and labor and capital as dependent variables instead of using the TFP measure. Although there are some studies that follow this specification (e.g., Badinger and Egger, 2016), the import share is still used for the foreign R\&D variable.} To augment CH's approach, some studies point out that the foreign R\&D variable used in a lot of studies captures only the current-period bilateral trade, and that indirectly generated demand must also be taken into account (e.g., Lumenga-Neso et al., 2005; Franco et al., 2011; Nishioka and Ripoll, 2012; Poetzsch, 2017). For example, it is possible that country A benefits from country C, D, and E's technology without importing from these sources, if country C, D, and E export to country B, which, in turn, exports to country A. This analysis uses the idea of Leontief inverse matrix to account for indirect demands, and suggests that these indirect trade-related spillovers are as important as the direct ones. It also strengthens the view that trade does matter for the international transmission of R\&D. Therefore, the introduction of indirect foreign R\&D spillovers into CH's approach seems to have resolved the doubts raised by Keller (1998) suggested the possibility that it does not matter from which a country imports on CH's approach.\\
\hspace*{3mm}There have been some studies on the effects of participation in the GVCs since the development of the global input-output table (e.g., Piermartini and Rubinova, 2014; Badinger and Egger 2016; Foster-McGregor et al., 2016; Poschl et al., 2016; Tajoli and Felice, 2018). However, these studies (including the above-mentioned studies) mostly investigate the overall use-driven R\&D spillover effects, and do not answer the question of which countries are important to trade with. In other words, these studies only examine the overall effects from abroad. In an attempt to answer this question, Keller (2002a) investigates the degree of localization of technology, using the distance between the G5 and other countries. He finds that for every additional 1200-km from the G5 countries there is a 50\% drop in technology diffusion. In this study, I analyze the geographical aspect, in terms of how the countries are connected, rather than the distance between them. A trade structure has long been confirmed that many countries trade with the hub in their region, and the hub trades with the hubs in other regions. Therefore, I classify the export side by ``centrality'', and understand the countries with whom it is important to.

\section{Data, Variables, and Classifications by Centrality}\label{sec3}
\subsection{Data}
\hspace{3mm}This study uses data from 14 manufacturing sectors and 21 countries\footnote{\;The countries are Belgium (BEL), Canada (CAN), Czech Republic (CZE), France (FRA), Germany (DEU), Greece (GRC), Hungary (HUN), Ireland (IRL), Italy (ITA), Japan (JPN), Mexico (MEX), Netherlands (NLD), Poland (POL), Portugal (PRT), Romania (ROM), Russia (RUS), Spain (ESP), Slovenia (SVN), Turkey (TUR), the United Kingdom (GBR), and the United States of America (USA).} for the years between 1995-2007 for the main analysis. This is because the business R\&D data at the industry level can be obtained only for 21 countries under the conditions of the interpolation of unreported data.\footnote{\;Data on R\&D expenditures is also available for Korea and Australia, however, I exclude them from my sample in this analysis. Japan is included in the data set because it is a G5 country, and its R\&D expenditures are considerably high. However, Korea and Australia are far from the other countries in the region, and their R\&D expenditure is not very high, therefore, their influence on the results is considered to be small. In addition, these two countries are excluded from the data set because including them in the centrality classification (see Section 4.2.3 for details) may cause bias in the results.} The global financial crisis of 2008 had a considerable impact on trade, therefore, including that year in the analysis is likely to bias the results. However, since I cannot collect much data after 2009, and the oldest global input-output table available dates to 1995, I test my hypothesis with data from 1995 to 2007.\\
\hspace*{3mm}Internationally comparable figures on employment, capital stocks, labor compensation, output, and value-added on the industry level come from the WIOD Socio Economic Accounts (WIOD SEA) for the years 1995-2007. To convert the nominal values to the real values, I use the industry-level value-added deflators, GDP deflators,\footnote{\;I use GDP deflator for only four countries (Ireland, Russia, Romania, and Turkey) because these countries are not reported in the STAN database.}, and constant 1995 dollars at purchasing power parity (PPP) rates. The value-added deflators and the GDP deflators are obtained from the OECD STAN database, and the World Bank's World Development Indicators, respectively. Data on sample countries' business R\&D come from the OECD ANBERD database, and the global input-output linkages data are obtained from the WIOD for the years 1995-2007.\\
\hspace*{3mm}An advantage of using the WIOD is that it provides socio-economic accounts, which include annual data such as employment and capital at the industry level. The OECD-ICIO is also available as an international input-output table, however, if this is used, it will be necessary to separately collect the variables used in the calculation of the TFP. It will also be necessary to convert them from nominal to real values.

\subsection{Definition of the variables}
\subsubsection{Industry-level TFP}
\hspace{3mm}I follow Caves et al. (1982) for constructing the industry-level TFP indexes. It is defined as:
\begin{align*}
\ln \mathrm{TFP}_{iht} = &\Bigl[\ln Y_{iht}-\frac{1}{N}\sum_{i}\ln Y_{iht}\Bigr] - \frac{1}{2}\Bigl(\sigma_{iht} + \frac{1}{N}\sum_{i}\sigma_{iht}\Bigr)\Bigl[\ln L_{iht} - \frac{1}{N}\sum_{i}\ln L_{iht}\Bigr]\\ &- \Bigl[1-\frac{1}{2}\Bigl(\sigma_{iht} + \frac{1}{N}\sum_{i}\sigma_{iht}\Bigr)\Bigr] \Bigl[\ln K_{iht} - \frac{1}{N}\sum_{i}\ln K_{iht}\Bigr],
\end{align*}
where $i$ is the country index, $h$ is the industry index, and $t$ is the time index. $Y$ is the real value-added, $\sigma$ is the labor compensation shares, $L$ is the total employment, $K$ is the real capital stock, and $N = 21$ is the number of countries.
Following Hall (1990), I use cost-based factors for labor shares since these are more robust than the revenue-based factor shares in the absence of a constant return to scale. Labor compensation shares are calculated as nominal labor compensation divided by nominal value-added. However, some of the obtained values of the labor shares exceed one.\footnote{\;This is because the value of capital compensation is negative.}
To control this, I use the smoothing procedure according to Harrigan (1997) such that the labor shares are obtained by regressing the following equation:
$$
\sigma_{iht} = \alpha_{i} + \alpha_{h} + \alpha_{t} + \phi_{h} \ln (K_{iht}/L_{iht}),
$$
where $ \alpha_{i}, \alpha_{h},$, and $\alpha_{t}$ are country, industry, and year fixed effects, respectively. I use the point estimates as the labor cost shares and the normalization of other variables for constructing the TFP, which should help reduce simultaneity issues (Keller 2002b).\\

\begin{table}[h]
\begin{center}
\begin{minipage}{\textwidth}
\caption{Summary statistics of TFP}\label{tab1}
\begin{tabular*}{\textwidth}{@{\extracolsep{\fill}}lccccc@{\extracolsep{\fill}}}
\toprule
& \multicolumn{5}{@{}c@{}}{All observations}\\ \cmidrule{2-6}
&Obs. & Min & Max & Mean & Std. dev. \\
\midrule
Food, Beverages, and Tobacco & 273 & -0.55 & 0.44 & -0.01 & 0.17 \\
Textiles and Textile Products & 273 & -0.53 & 0.37 & -0.01 & 0.21 \\
Leather, Leather and Footwear & 273 & -0.7 & 0.55 & -0.01 & 0.25 \\
Wood and Products of Wood and Cork & 273 & -0.5 & 0.39 & -0.01 & 0.18 \\
Pulp, Paper, Printing, and Publishing & 273 & -0.32 & 0.64 & -0.01 & 0.18\\
Coke, Refined Petroleum, and Nuclear Fuel & 273 & -1.88 & 0.94 & -0.02 & 0.47 \\
Chemicals and Chemical Products & 273 & -0.56 & 0.75 & -0.01 & 0.24 \\
Rubber and Plastics & 273 & -0.35 & 0.28 & -0.01 & 0.15 \\
Other Non-Metallic Mineral & 273 & -0.47 & 0.24 & -0.01 & 0.14  \\
Basic Metals and Fabricated Metal & 273 & -0.4 & 0.28 & -0.01 & 0.15\\
Machinery, Nec & 273 & -0.73 & 0.32 & -0.01 & 0.23 \\
Electrical and Optical Equipment & 273 & -0.58 & 0.58 & -0.01 & 0.2 \\
Transport Equipment & 273 & -0.47 & 0.33 & -0.01 & 0.18\\
Manufacturing, Nec; Recycling & 273 & -0.38 & 0.52 & -0.01 & 0.16\\
\toprule
\multicolumn{5}{l}{Note : Author's calculation. This table is constructed using all observations.}
\end{tabular*}
\end{minipage}
\end{center}
\end{table}

\hspace*{3mm}Table 1 reports, for each industry, the summary statistics for the TFP for all the observations and the growth rate obtained by dividing the value for 2007 by the value for 1995. Across all the samples, the mean is almost the same for all the industries, except for ``Coke, Refined Petroleum, and Nuclear Fuel'' where the maximum value is higher than the others while the minimum value is much smaller, indicating that there is a large difference in the technical competence in that industry.

\subsubsection{Industry-level R\&D stocks}
\hspace*{3mm}Next, I explain the computation of domestic R\&D stock. In this study, I use only the manufacturing R\&D, and I set non-manufacturing R\&D as zero.\footnote{\;Non-manufacturing sectors' R\&D are quite noisy, and there are a lot of unreported values.} The domestic R\&D stock $S_{iht}$ for industry $h$ in the country $i$ at the time $t$ is computed using the perpetual inventory methods as follows:
$$
S_{iht} = (1-\delta)S_{ih,t-1} + R_{iht},
$$
where $\delta$ is the depreciation rate of knowledge obsolescence, assumed to be 0.15 following Coe and Helpman (1995), and $R_{iht}$ is the real business R\&D expenditure. To obtain the initial value of the real business R\&D stock, I compute:
$$
S_{ih0} = \frac{R_{ih0}}{\delta + g_{ih}},
$$\\
where $g_{ih}$ is the annual average logarithmic growth rate of real business R\&D expenditures for industry $h$ in the country $i$ for the entire period for which data are available. Real business R\&D expenditures are obtained from the OECD ANBERD database; however, the values are deflated by country-level GDP deflators. To reduce the possibility of results being induced by common deflators, I use the value-added deflators at the industry level reported in the STAN database.\\
\hspace*{3mm} Using domestic R\&D stock and the WIOD, I calculate the R\&D contents of the final goods in equation (1) for 40 countries and RoW.\footnote{\;The other countries, in addition to the previously-mentioned 21 countries, were Australia (AUS), Austria (AUT), Bulgaria (BGR), Brazil (BRA), China (CHN), Cyprus (CYP), Denmark (DNK), Estonia (EST), Finland (FIN), India (IND), Indonesia (IDN), Korea (KOR), Latvia (LVA), Lithuania (LTU), Luxembourg (LUX), Malta (MLT), Slovakia (SVK), Sweden (SWE), and Taiwan (TWN) } After calculation, I selected the 21 countries since I could not obtain the domestic R\&D contents for the other countries.

\subsubsection{The R\&D contents of the final goods}
\hspace*{3mm}Following Nishioka and Ripoll (2012), I introduce the R\&D content of the ``final goods,'' which focuses on the R\&D embodied in trade volume, and considers both direct and indirect input flows. I first explain the definition of R\&D contents for the ``intermediates'' used, and proceed to show how to change it to the final goods.\\
\hspace*{3mm}Nishiokan and Ripoll (2012) follow Trefler and Zhu (2010), and use a diagonal matrix, $Q$, of dimension $NG \times NG$ where the element $Q_{i}(h)$ is the amount of industry $h$'s output in the country $i$. They also use a global intermediate input coefficients matrix, $B$, of dimension $NG \times NG$ where the element $B_{ji}(g,h)$ is the amount of intermediate input $g$ made in the country $j$ to produce one unit of gross output of country $i$'s good $h$. If $a_{ji}(g,h)$ is the value of goods shipped from industry $g$ in country $j$ for intermediate use by industry $h$ in country $i$, $B_{ji}(g,h)$ is calculated by $a_{ji}(g,h) / Q_{i}(h)$.
Now consider the demand of intermediate inputs, including indirect effects, using these matrixes. Let $Z_{i}$ be a $NG \times 1$ vector with the $i$th element representing global consumption of goods from country $i$, while all the remaining elements are zero. The production of $Z_{i}$ requires intermediate inputs given by $BZ_{i}$. Further, since the production of these intermediate inputs requires the use of other intermediate inputs given by $B^{2}Z_{i}$, if I consider it as a recursive algorithm, the matrix of intermediate inputs, directly and indirectly, required for the production of $Z_{i}$ is given by $\sum_{n=1}^{\infty} B^{n}Z_{i}$. Thus, the total requirements needed to deliver $Z_{i}$ is given by
$$
Z_{i} + \sum_{n=1}^{\infty} B^{n}Z_{i} = \sum_{n=0}^{\infty} B^{n}Z_{i} = (I-B)^{-1}Z_{i},
$$
where $I$ is the identity matrix, and $(I-B)^{-1}$ is known as the Leontief inverse matrix. This equation allows me to calculate the direct and indirect effects on the demand for intermediate goods. Nishioka and Ripoll (2012) use $BQ$ instead of $Z_{i}$ in their analysis. Next, let $S_{j}$ be a $1 \times G$ vector whose $g$th element is the real business R\&D stock to produce $g$ in country $j$, and let $D_{j}$ be a $1 \times G$ vector whose $g$th element is the ratio of the R\&D stock to good $g$ in country $j$. Thus, the direct R\&D requirements are defined as $D_{j}Q_{j} = S_{j}$. Nishioka and Ripoll (2012) define the R\&D content of intermediate inputs $F$ as:
$$
F = D(I-B)^{-1}BQ.
$$
\hspace{3mm}They also define $(I-B)^{-1}B = A$, and select the elements of $D$ and $A$ for capturing the various effects. For example, to calculate the total (domestic and foreign) effects of trading with the same industries, I compute $\sum_{j}D_{j}(h)A_{ji}(h,h)Q_{i}(h)$, and if the trades are between different industries, I calculate $\sum_{j}\sum_{g\neq h}D_{j}(g)A_{ji}(g,h)Q_{i}(h)$. Using these disaggregations, they analyze similar investigation as Keller (2002b) in the years 1995, 2000, and 2005. They find that the R\&D knowledge spillovers through trade in intermediate inputs are only among high R\&D industries, and it is statistically significant for trade between the same industries, however, it is insignificant for trade between different industries. Thus, their results suggest that the effects of R\&D on productivity with intermediate trade are limited, and these points are different from Keller's (2002b).\\
\hspace*{3mm}In this study, I modify their model. Although Nishioka and Ripoll (2012) use the intermediate goods matrix ($BQ$) for $Z_{i}$, the methodology of Trefler and Zhu (2010) correspond to the equilibrium output model in the input-output analysis. Therefore, the final demand matrix should be used for $Z_{i}$. In the first place, the number of columns is different between the two ($BQ$ and $Z_{i}$). Therefore, I use the final demand and calculate the R\&D contents embodied in the final demand in my analysis.\\ 
\hspace*{3mm}To reconstruct the R\&D content of the final goods, I use a diagonal matrix of the sum of exports to global final demand of country $j$, $f_{j}$, of dimension $G \times G$:
$$
f \equiv \left[
    \begin{array}{cccc}
      f_{1} & 0 & \ldots & 0 \\
      0 & f_{2} & \ldots &0 \\
      \vdots & \vdots & \ddots & \vdots \\
      0 & 0 & \ldots & f_{N}
    \end{array}
  \right].
$$\\
Thus, the R\&D content of the final goods for the final demand is defined as follows:
\begin{eqnarray}
V = D(I-B)^{-1}f.
\end{eqnarray}
\subsection{Classifications by Centrality}
\subsubsection{Centrality in production networks}
\hspace*{3mm}The concept of centrality is originally used in sociology, and various centrality measures have been used to capture a person's centrality, power, prestige, and influence. Among several measures of centrality, those used to identify the influence of individuals within an organization are called the ``Degree centrality'' and the ``Eigenvector centrality,'' as defined by Bonacich (1972), and the ``Katz-Bonacich eigenvector centrality,'' as defined by Katz (1953) and Bonacich (1987). In these measures, high centrality indicates that a person is important to the organization. These centrality measures have been used in sociology to identify individuals who have a significant influence on their surroundings within organizations.\footnote{\;For example, Alatas et al. (2016) examine the influence of network structure on information aggregation, using the social structure of Indonesian villages, and find that the presence of central households with an easy access to information and connections with others play an important role in increasing the efficiency of knowledge diffusion. Additionally, Acemoglu et al. (2014) show that overall learning efficiency is achieved by the presence of a person who is the hub of the information.}\\
\hspace*{3mm}Using this idea, a similar analysis can be performed for economics by regarding the organization as a production network in economic activity, and the individuals as firms, industries, or countries.
In fact, the existence of hubs in production network is also confirmed through data\footnote{\;Bernard et al. (2007) identified the existence of hubs at the firm level, using customs data, and Criscuolo and Timmis (2018a) identified the same at the industry and country level, using global input-output table.}, and research regarding the importance of hubs has also been conducted, using these centrality measure. For example, the study of propagation of shocks in macroeconomics used the idea of centrality (Acemoglu et al., 2012; Carvalho, 2014; Acemoglu et al., 2016; Boehm et al., 2019; Carvalho et al., 2021). These studies used firm-level data. However, Criscuolo and Timmis (2018a) used the global input-output table to conduct macro-level analysis. They examined the change of centrality in a country and an industry both in backward and forward linkages. They found that in 1995, a small number of central hubs, such as Japan, Germany, and the United States, dominated each regional value chain, and that in 2011, the central hub of Asia changed from Japan to China. In addition, Criscuolo and Timmis (2018b) examined the effects of the changes in the GVC structures on firm productivity, taking into account centrality in production networks. They found that increasing centrality was important for improving the productivity of smaller or non-frontier firms, and this trend has been witnessed post-2004 in firms in the EU-member or small countries.\\
\hspace*{3mm}I have discussed the importance of the presence of central countries, industries, and firms in a production network. However, there is no study on the existence of hubs for the diffusion of knowledge through trade. Currently, most research are regarding the propagation of shocks. Previous studies of international R\&D spillover effects have focused on direct and indirect relationships, however, no analysis has as yet incorporated the idea of centrality into their study.
\begin{figure}[h]
  \begin{center}
     \includegraphics[width=0.9\textwidth]{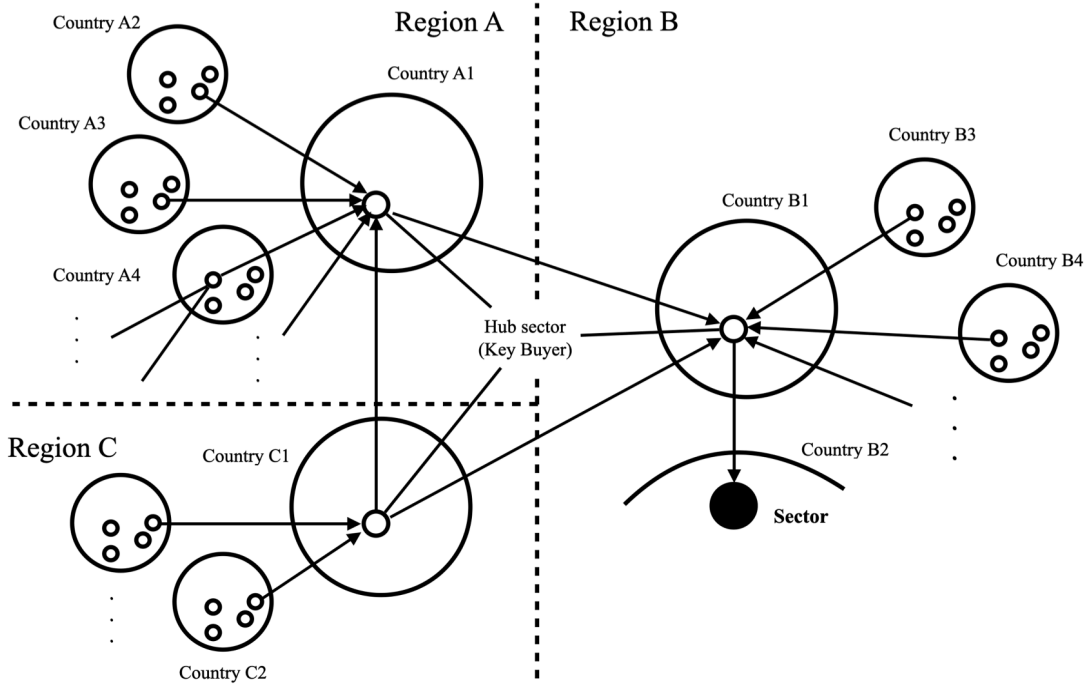}
   \end{center}
\caption{Overview of trade network}
\end{figure}
I consider that the central countries in GVCs are connected to various countries and, thus, are important as mediators in gathering the technologies in trade from other countries. For example, Figure 1 shows an overview of an industry in Country B2 that imports the final goods from Country B1. Country B1 imports intermediate goods not only from the same region, but also from other regions. The hubs in other regions (Country A1 and Country C1) also import intermediate goods from those regions, and export to Country B1. Therefore, the sector in Country B2 receives the knowledge of the countries in other regions via the hub sector, Country B1, in the same region. In other words, I consider that the central countries in the GVCs have large spillovers in the surrounding countries through large indirect demands. Figure 1 shows the effect of the countries that are central to the ``imports,'' however, I am interested in the spillover effects of the countries that are also central to knowledge diffusion, or ``exports.'' By estimating the effects of the countries that are central to both imports and exports, I confirm the effect of knowledge gathering, and the extent of its diffusion. Therefore, in the next subsection, I identify the centrality of the countries and the industries, using the centrality measure.
\subsubsection{Centrality Measures}
\hspace{3mm}In this study, I use the Katz-Bonacich eigenvector centrality.\footnote{This measure has been widely used in literature refarding the propagation of shocks.} This measure can rank the influence of each node to evaluate its importance in more detail. For example, Figure 1 shows that a simple production network and the size of each node denotes centrality. It is worth noting here that A has the highest centrality, not C. B and C exist as a hub for G and H, and D, E, and F, respectively, and A exists as a hub that connects B and C. In the Katz-Bonacich eigenvector centrality, it is important whether a shock is likely to propagate significantly across the whole, not simply the number of connections with others.
\begin{figure}[h]
  \begin{center}
     \includegraphics[width=0.63\textwidth]{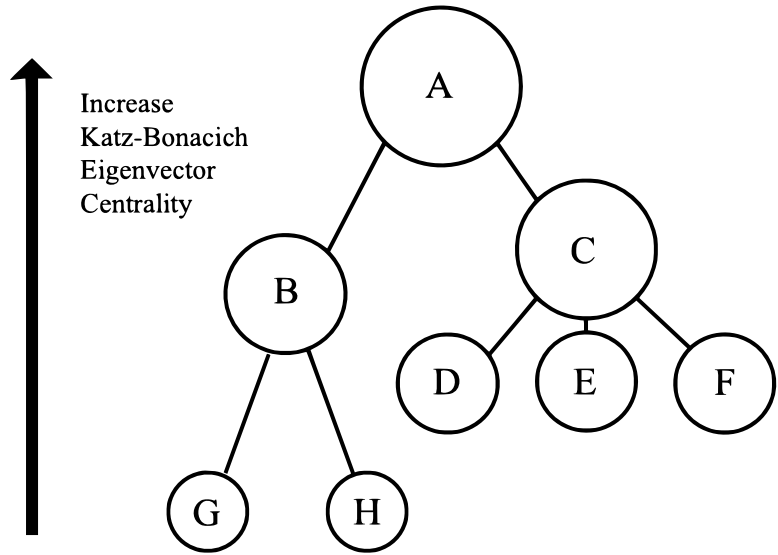}
   \end{center}
\caption{Katz-Bonacich eigenvector centrality}
\footnotesize{Source: Criscuolo and Timmis(2018a)}\\
\footnotesize{Note: The size of each node denotes centrality. This measure of centrality represents a node as a central hub if it can easily influence the whole. Thus, A, the hub of B and C, is the central hub.}
\end{figure}
As discussed above, other metrics, i.e, ``degree centrality'' and ``eigenvector centrality'' are popular in sociology, however, there are reasons why they are not suitable for this study.
First, the degree centrality is defined as the sum of the number of outdegree or indegree,\footnote{\;Outdegree represents the edges directed out from a vertex in a network, while indegree represents the edges directed toward a vertex. In trade, we can think of the vertex as a country or an industry.} however, it captures only direct connections. This measure results in several industries with the same centrality in trade, making it difficult to group them by the degree of their centrality. 
Second, although the eigenvector centrality captures both direct and indirect effects, however, if either the number of outdegree or indegree is zero, then the centrality becomes zero. In trade, since several industries only import or export,\footnote{\;This problem is seen in all sectors: agriculture, manufacturing, and services.} this measure is not suitable for this analysis.\\
\hspace*{3mm}On the other hand, the Katz-Bonacich eigenvector centrality solves the above-mentioned problems. However, this measure denotes the nodes that trade more with the central node also as important nodes, even if the total trade value of the node is smaller than the others. Nevertheless, this is not a concern in this analysis because those nodes (i.e., countries or industries) can be considered important transmitters of the central node's knowledge.\\
\hspace*{3mm}By using the world input-output table, I compute the centrality of each sector in each country. It is important to note that a centrality obtained by calculation does not represent centrality within the same industry alone, but rather all the other industries. Furthermore, since the effects on productivity of trading with central industries in exports and imports, respectively, are likely to be different, I distinguish between forward linkages and backward linkages in my analysis. Following Criscuolo and Timmis (2018a), I calculate forward ($\mathrm{Centrality}_{iht}^{fwd}$) and backward ($\mathrm{Centrality}_{iht}^{back}$) centrality that are defined as:
\begin{eqnarray*}
\mathrm{Centrality}_{iht}^{fwd} &=& \eta (\mathbf{I} - \lambda \mathbf{W})^{-1} \mathbf{1},\\
\mathrm{Centrality}_{iht}^{back} &=& \eta (\mathbf{I} - \lambda \mathbf{W^\top})^{-1} \mathbf{1},
\end{eqnarray*}
where $W$ represents input-output linkages matrix, and $\eta = 1- \lambda$ is the baseline centrality level. The parameter, $\lambda$, reflects the average share of intermediate inputs in production, as shown theoretically by Acemoglu et al. (2012) and Carvalho (2014). I use $\lambda = 0.5$.
When I calculate forward and backward centrality, I assume $\sum_{i}w_{ji}=1$ and $\sum_{j}w_{ji}=1$, respectively.
If $a_{ji}(g,h)$ is the value of goods shipped from sector $g$ in country $j$ for intermediate use by sector $h$ in country $i$, the elements of matrix $W$ are defined as $w_{ji}(g,h) = a_{ji}(g,h) / \sum_{i}a_{ji}(g,h)$ in forward linkages, while $w_{ji}(g,h) = a_{ji}(g,h) / \sum_{j}a_{ji}(g,h)$ in backward linkages.\\
\hspace*{3mm}There are two things to note in the measurement of centrality used in this study. First, when I calculate centrality, I exclude RoW in the WIOD (i.e., I use 40 countries). The Katz-Bonacich eigenvector centrality reflects that nodes with relationships with larger nodes are also large. Since the RoW is a collection of various countries, it can be a large node. Therefore, due to the existence of a huge node called RoW, the centrality of the other countries may not be reflected well. For example, even if a country is not central by nature, but conducts a lot of trade with the RoW, its centrality will increase.
\hspace*{3mm}Second, I also exclude the internal transactions in a country from the input-output table. In this study, I aim to investigate the impact of the industries that are in important positions of the GVCs. If I include the internal transactions of a country in the calculation, a country's centrality in the GVCs will be higher if the internal transactions are considerably more active than those with the other countries. However, the high centrality industries are not considered important in the GVCs, and it deviates from the objective of this study. Thus, I exclude the internal transactions from the input-output table. These notations indicate that the higher the value of the forward centrality, the greater the supply of the industry to similar and other industries in trade. Additionally, the higher the value of the backward centrality, the greater the demand from the industry from similar and other industries in trade. Therefore, it can be said that industries with high centrality exist as the hubs in trade. Under these notations, after the calculation of centrality for 40 countries, I selected 21 countries.\\
\hspace*{3mm}Table 2 displays the summary statistics for backward and forward centrality for each industry in 21 countries. Table 2 shows that the maximum value of centrality is so difference among industries. As discussed above, the degree of centrality represents the importance in trade. For example, the maximum value of backward centrality for Transport Equipment is 11.26, which is quite high, and the country with this centrality imports from many countries and industries when the final demand per unit occurs. The maximum value of forward centrality for Chemicals and Chemical products is also quite high at 18.77. The country with this centrality supplies the most when the final demand for each industry occurs. For industries with small maximum values, the values of centrality of industries are generally small, indicating that they are not important in trade. On the other hand, for industries with large maximum values, it is considered that there are countries that are hubs in trade, i.e., countries with high centrality, such as Transport Equipment and Chemicals and Chemical products. The maximum values of forward centrality within industry are larger than that of backward centrality, but this may be because exports are concentrated in some countries because exports have higher hurdles than imports (Bernard and Jensen, 1999; Clerides et al., 1998; Melitz, 2003; Helpman et al, 2004).

\begin{table}\centering
\caption{Summary statistics of backward and forward centrality}\label{tab2}
\tiny
\begin{tabular*}{\textwidth}{@{\extracolsep{\fill}}lccccccccccc@{\extracolsep{\fill}}}
\toprule
& \multicolumn{5}{@{}c@{}}{Backward centrality} & \multicolumn{5}{@{}c@{}}{Forward centrality}\\ \cmidrule{2-11}
 & Obs. & Min & Max & Mean & Std. dev &   Obs. & Min & Max & Mean & Std. dev \\
 \toprule 
Food, Beverages, and Tobacco & 273 & 0.59 & 5.13 & 1.89 & 1.13  &273 & 0.51 & 2.21 & 0.89 & 0.48 \\
Textiles and Textile Products & 273 & 0.55 & 3.15 & 1.16 & 0.62 &  273 & 0.51 & 4.60 & 1.22 & 0.87 \\
Leather, Leather and Footwear & 273 & 0.51 & 2.41 & 0.74 & 0.34   &273 & 0.51 & 2.55 & 0.69 & 0.40 \\
Wood and Products of Wood and Cork & 273 & 0.53 & 1.84 & 0.79 & 0.32  &273 & 0.52 & 2.25 & 0.86 & 0.37 \\
Pulp, Paper, Printing, and Publishing & 273& 0.56 & 3.97 & 1.21 & 0.76 &   273 & 0.52 & 6.02 & 1.58 & 1.33 \\
Coke, Refined Petroleum, and Nuclear Fuel & 273 & 0.51 & 2.00 & 0.86 & 0.35 &273 &0.52 & 5.70 & 1.34 & 1.04 \\
Chemicals and Chemical Products & 273 & 0.57 & 4.97 & 1.46 & 1.02 &  273 &0.64 & 18.77 & 4.14 & 4.16 \\
Rubber and Plastics & 273 & 0.55 & 2.32 & 0.88 & 0.41 &  273 & 0.51 & 3.16 & 1.28 & 0.72 \\
Other Non-Metallic Mineral & 273 & 0.53 & 2.11 & 0.81 & 0.29 &  273 & 0.52 & 2.10 & 0.87 & 0.40 \\
Basic Metals and Fabricated Metal & 273 & 0.59 & 5.57 & 1.54 & 1.10 &  273 & 0.61 & 12.9 & 2.77 & 2.46 \\
Machinery, Nec & 273 & 0.56 & 5.37 & 1.25 & 1.03 &  273 & 0.55 & 8.97 & 1.86 & 1.96 \\
Electrical and Optical Equipment & 273 & 0.56 & 5.19 & 1.58 & 1.15 & 273 & 0.50 & 12.6 & 2.86 & 3.06 \\
Transport Equipment & 273 & 0.54 & 11.26 & 2.42 & 2.49 &  273 &0.50 & 10.07 & 2.14 & 2.22 \\
Manufacturing, Nec; Recycling & 273 & 0.53 & 2.55 & 0.96 & 0.47 & 273 & 0.51 & 2.06 & 0.82 & 0.35\\
\toprule
\multicolumn{5}{l}{Note : Author's calculation. This table is constructed using all observations.}
\end{tabular*}
\end{table}

\subsubsection{Classification of exporters}
\hspace*{3mm}To explore which country has the high centrality, I divided the industries into three categories: high, middle, and low. I defined the top 33\% as high centrality, the next top 33\% as middle centrality, and the remaining as low centrality in each sector, in each year, respectively.\\
\hspace*{3mm}In this study, I am interested in countries that have both a ``gathering'' and ``diffusing'' nature of knowledge, therefore, I further classify the sample countries by total centrality that was defined by using both backward and forward centrality. Countries that are ``high'' in both backward and forward linkages are considered ``High'', those that are ``low'' in both are considered ``Low'', and the remaining are categorized as ``Middle. ''\footnote{\;Although Criscuolo and Timmis (2018a) define total centrality as the average of the backward and forward centrality, I categorize the countries in this way because only one of them may have a large value, and even if the total centrality is large, it may not necessarily have the properties of both knowledge aggregation and diffusion.}\footnote{\;There are a total of seven combinations of Middle: high (in backward)-middle (in forward); high-low; middle-high; middle-middle; middle-low; low-high; low-middle.} High, Middle, and Low account for about 27\%, 47\%, and 26\% of the total, respectively.\footnote{\;In Middle's seven categories, high-middle accounts for 11.4\%, high-low for 1.8\%, middle-high for 12.6\%, middle-middle for 45.9\%, middle-low for 13.2\%, low-high for 0.7\%, and low-middle for 14.4\%, on an average, from 1995 to 2007.}\\
\hspace*{3mm}I divide a country into which classification it belongs in each industry, but also divide the industry itself into two groups. This is because, as Table 2 shows, there is a clear difference in the maximum value of centrality across industries. 
For example, the maximum value for ``Wood and Products of Wood and Cork'' is 1.84 and 2.25 in backward and forward linkages, respectively, while for ``Transport equipment'' it is 11.26 and 10.07, respectively. Since, centrality is computed as a relative value across all the trade, therefore, even if a country is defined as a high group in the ``Wood and Products of Wood and Cork'' category, it cannot be called a hub in the overall trade. Based on this, I classify the top seven industries with maximum values of centrality as ``Group B'', and the rest as ``Group A''.
\section{Empirical Strategy}\label{sec4}
\subsection{Basic model}
\hspace{3mm}Following Nishioka and Ripoll (2012), the benchmark model used in the literature to investigate the spillover effects of R\&D on industry productivity through trade in the final goods is:
\begin{align}
\ln \mathrm{TFP}_{iht} = \alpha_{ih} + \alpha_{t} + \beta \ln V_{iht} + \epsilon_{iht},
\end{align}
where indexes $i$, $h$, and $t$ denote country, industry, and year, respectively. $\mathrm{TFP}_{iht}$ measures the productivity of country $i$ in industry $h$ in the year $t$. $\alpha_{ih}$ and $\alpha_{t}$ are country-industry and time-fixed effects, respectively. $V$ is the R\&D content of the final goods in the for final demand, and this is defined in equation (1). $\epsilon_{iht}$ is the error term. In this equation, I use the R\&D content of the final goods, $V$, instead of $F$, as defined by Nishioka and Ripoll (2012). Following Bertrand et al. (2004), I address the potential serial correlation and heteroskedasticity issues by calculating the standard errors clustered at the industry level. I avoid the simultaneity issues in two ways. First, I employ the TFP index of Caves et al. (1982). Although this method requires more variables, and is more difficult to construct than the method of Coe and Helpman (1995) and Coe et al. (2009), it reduces the simultaneity problems. Second, I use industry-specific value-added deflators for almost all the countries. The real R\&D expenditures reported in the OECD ANBERD are deflated by the GDP deflators, however, the impact of the common deflator is reduced by using industry-level value-added deflators.\\
\hspace*{3mm}The analysis in this study deepens the study of the effects of R\&D on productivity through trade by incorporating data that was excluded from previous studies, and by considering new ideas of centrality that further elucidate the structure of the GVCs.
\subsection{Model with classification by centrality}
\hspace*{3mm}In the empirical analysis, I decompose equation (2) into domestic and foreign R\&D inputs, and analyze the spillover effects from foreign R\&D contents, including centrality.\\
\hspace*{3mm}First, I use the two industry groups, and the three categories (High, Middle, and Low) for centrality, as described above, to investigate whether the centrality of the exporting country affects the spillover effects, both in more centralized industries and others. Since countries with high centrality have strong relationships with various countries, they are expected to produce larger effects than countries with middle or low centrality. I also focus on the importing country's region in this analysis. This is because there are more hubs in Europe than in the other regions, and European countries are more likely to be involved with other hub countries, therefore, a larger effect is expected. To observe this effect, I divided the sample on the import side into three regions: All (ALL), Europe (EUR), and North America (NA).\footnote{\;I don't separate sample in Asia because Japan is the only Asian country in my sample.}\\
\hspace*{3mm}For the importers to I make ALL, EUR, and NA sets of all the sampled countries, only European countries in the sample, and only North American countries in the sample, respectively, and make A and B sets of Group A and Group B, respectively. For the exporters, I also make High, Middle, and Low sets of industries included in the High, Middle, and Low centrality categories, respectively. I estimate the following specifications:
\begin{align}
\ln \mathrm{TFP}_{iht}=\:\:&\alpha_{ih} + \alpha_{t} + \beta \ln \Bigl[\sum_{g}D_{it}(g)T_{iit}(g,h)\Bigr]\nonumber\\& 
+\rho_{1} \ln \Bigl[\sum_{j \neq i, i \in \mathrm{ALL}\;\mathrm{or}\;\mathrm{EUR}\;\mathrm{or}\;\mathrm{NA},\;j \in \rm{High}}\sum_{\;g \in \;\mathrm{A}\;\mathrm{or}\;\mathrm{B}}D_{jt}(g)T_{jit}(g,h)\Bigr]\nonumber\\& 
+ \rho_{2} \ln \Bigl[\sum_{j \neq i, i \in \mathrm{ALL}\;\mathrm{or}\;\mathrm{EUR}\;\mathrm{or}\;\mathrm{NA},\;j \in \rm{Middle}}\sum_{\;g \in \;\mathrm{A}\;\mathrm{or}\;\mathrm{B}}D_{jt}(g)T_{jit}(g,h)\Bigr] \nonumber\\&
+ \rho_{3} \ln \Bigl[\sum_{j \neq i, i \in \mathrm{ALL}\;\mathrm{or}\;\mathrm{EUR}\;\mathrm{or}\;\mathrm{NA},\;j \in \rm{Low}}\sum_{\;g \in \;\mathrm{A}\;\mathrm{or}\;\mathrm{B}}D_{jt}(g)T_{jit}(g,h)\Bigr] + \epsilon_{iht},
\end{align}
where the first term in brackets ($\beta$) captures the effect of domestic R\&D contents, and the second to fourth terms ($\rho_{1}, \:\rho_{2}, \: \mathrm{and} \: \rho_{3}$) capture the effect of the foreign R\&D contents. Additionally, $j$ and $g$ are subscripts, indicating the exporter.
\subsection{Comparison of the effects between G5 and non-G5 countries}
\hspace*{3mm}Next, I estimate the effect, which classifies importers on technological aspects instead of geographical aspects, like equation (3). I categorize importers into G5 and non-G5. The share of domestic and foreign R\&D contents in all the sample countries indicates that there is a large difference between G5 and non-G5. For example, the share of domestic R\&D contents in Japan is about 96-97.5\%, the United States is about 92-93\%, and Germany, France, and the U.K. are about 70-80\%. The share of foreign R\&D contents in Germany, France, and the U.K. is relatively high compared to Japan and the United States, probably because they are in the EU. Nevertheless, the gap with non-G5 countries is large. Therefore, as an assumption, it is expected to be affected from within in the case of high technology countries. I estimate the specification as follows: 
\begin{align}
\ln \mathrm{TFP}_{iht}=\:\:&\alpha_{ih} + \alpha_{t}+ \beta \ln \Bigl[\sum_{g}D_{it}(g)T_{iit}(g,h)\Bigr]\nonumber\\& + \upsilon_{1} \ln \Bigl[\sum_{j \neq i, i \in \mathrm{G5}\;\mathrm{or}\;\mathrm{nonG5},\;j \in \rm{High}}\sum_{\;g \in \mathrm{A}\;\mathrm{or}\;\mathrm{B}}D_{jt}(g)T_{jit}(g,h)\Bigr]\nonumber\\& 
+ \upsilon_{2} \ln \Bigl[\sum_{j \neq i, i \in \mathrm{G5}\;\mathrm{or}\;\mathrm{nonG5},\;j \in \rm{Middle}}\sum_{\;g \in \mathrm{A}\;\mathrm{or}\;\mathrm{B}}D_{jt}(g)T_{jit}(g,h)\Bigr] \nonumber\\&
+ \upsilon_{3} \ln \Bigl[\sum_{j \neq i, i \in \mathrm{G5}\;\mathrm{or}\;\mathrm{nonG5},\;j \in \rm{Low}}\sum_{\;g \in \mathrm{A}\;\mathrm{or}\;\mathrm{B}}D_{jt}(g)T_{jit}(g,h)\Bigr] + \epsilon_{iht}.
\end{align}
\section{Estimation Results}\label{sec5}
\subsection{Main results}
\hspace*{3mm}In this part, I explain the results of estimating equation (3), which classifies importers on geographical aspects, and considers the scale of the industries and exporters' centrality in trade.\\
\begin{table}[h]
\begin{center}
\begin{minipage}{\textwidth}
\caption{The results of exporters' centrality effects}
\begin{tabular*}{\textwidth}{@{\extracolsep{\fill}}lcccccc@{\extracolsep{\fill}}}
\toprule%
& \multicolumn{3}{@{}c@{}}{Group A} & \multicolumn{3}{@{}c@{}}{Group B} \\\cmidrule{2-7}%
 &        (1)         &        (2)          &        (3)         &        (4)      &        (5)          &        (6)        \\
            &        All         &        Europe          &        North America         &        All         &        Europe          &        North America \\
\midrule
domestic, $\beta$&-0.019&      -0.028         &      $0.045^{**}$        &      -0.025         &      -0.032         &       $0.045^{**}$  \\
            &     (0.015)         &     (0.019)         &     (0.023)         &     (0.015)         &     (0.020)         &     (0.021)         \\
\textbf{exporter's}\\
\textbf{centrality:}\\
\addlinespace
\hspace{3mm}\rm{High}, $\rho_{1}$   &       $0.027^{*}$ &       0.014         &       0.014         &       $0.040^{**}$         &       $0.063^{***}$ &       0.013         \\
            &     (0.015)         &     (0.021)         &     (0.028)         &     (0.019)         &     (0.019)         &     (0.054)         \\
\addlinespace
\hspace{3mm}\rm{Middle}, $\rho_{2}$   &    0.027         &       0.050         &       -0.031         &       $0.034^{**}$ &       $0.036^{**}$ &      -0.061         \\
            &     (0.019)         &     (0.028)         &     (0.028)         &     (0.014)         &     (0.015)         &     (0.040)         \\
\addlinespace
\hspace{3mm}\rm{Low}, $\rho_{3}$   &       0.015         &       0.010         &       0.004         &       $0.015^{*}$  &       0.008         &      -0.003         \\
            &     (0.012)         &     (0.016)         &     (0.017)        &     (0.008)         &     (0.011)         &     (0.025)          \\
\addlinespace
$\mathrm{constant}$&       0.105         &       0.132         &       -0.134 &      -$0.110^{*}$  &        			-$0.150^{*}$  &      -0.009         \\
            &     (0.096)         &     (0.123)         &     (0.159)         &     (0.061)         &     (0.070)         &     (0.241)         \\
\(N\)       &        3819         &        3091         &         546        &        3819         &        3091         &         546          \\
\(R^{2}\)   &       0.867         &       0.855         &       0.957        &       0.870         &       0.860         &       0.958          \\
\midrule
Country-\\
Industry FE   &        Yes         &        Yes         &        Yes         &        Yes         &        Yes         &        Yes                  \\
Year FE   &        Yes         &        Yes         &        Yes         &        Yes         &        Yes         &        Yes                 \\
\toprule
\end{tabular*}
\footnotetext{Note: This table shows the results for using the classification of centrality and the two industry groups ”All” includes 21 countries. Standard errors in parentheses, \footnotesize{* \(p<0.1\), ** \(p<0.05\), *** \(p<0.01\)}}
\end{minipage}
\end{center}
\end{table}
Columns (1) to (3) in Table 3 show the results for Group A, an industry with relatively low centrality, and columns (4) to (6) show the results for Group B, an industry with relatively high centrality. First, Table 3 shows that industries in Group A are hardly statistically significant. This is probably because these industries are less important and less active in trade, therefore, less susceptible to foreign influences. Second, in Group B for All and Europe, the impact of foreign R\&D contents from high centrality exporters is positively significant, and the largest. That is, trade with countries with high centrality generates the largest spillovers. Since these countries are defined as central to both imports and exports, the results suggest that they are mediators of knowledge gathering and diffusion. Third, the results show that countries with middle centrality are also important in the spillover effect. This result is discussed in detail in Section 5.3 (Discussion) and Appendix. Finally, in contrast to the other country groups, the result is positively significant for domestic R\&D contents in North America.

\subsection{Difference in the results between G5 and non-G5 countries}
\hspace*{3mm}In the previous subsection, Table 3 showed that the results for All samples and European countries were consistent with the hypothesis, while only North American countries obtained spillover effects from domestic R\&D contents. However, since North America has only three countries, it may strongly reflect the influence of the U.S., an economic superpower. Next, I describe the results of estimating equation (4), which classifies importers on technological aspects instead of geographical aspects, like equation (3).
\begin{table}[h]
\begin{center}
\begin{minipage}{\textwidth}
\caption{G5 and non-G5 results}
\begin{tabular*}{\textwidth}{@{\extracolsep{\fill}}lcccccc@{\extracolsep{\fill}}}
\toprule%
& \multicolumn{2}{@{}c@{}}{Group A} & \multicolumn{2}{@{}c@{}}{Group B} \\\cmidrule{2-5}%
 &        (1)         &        (2)          &        (3)         &        (4)           \\
            &        G5         &        Non-G5          &        G5         &        Non-G5        \\
\midrule
\rm{domestic}, $\beta$&       $0.102^{**}$ &      -0.021    &     $0.095^{***}$&      $$-$0.029^{*}$  \\
            &     (0.043)         &     (0.015)        &    (0.036)         &     (0.016)          \\
\textbf{exporter's}\\
\textbf{centrality:}\\
\addlinespace
\hspace{3mm}\rm{High}, $\upsilon_{1}$  &      -0.019         &       0.029            &  $0.058^{**}$          &       $0.060^{**}$  \\
            &     (0.023)         &     (0.021)       &    (0.021)         &     (0.020)           \\
\addlinespace
\hspace{3mm}\rm{Middle}, $\upsilon_{2}$ &       0.021  &       0.033      &      -0.013         &       $0.039^{**}$          \\
            &     (0.023)         &     (0.026)         &    (0.020)         &     (0.018)         \\
\addlinespace
\hspace{3mm}\rm{Low}, $\upsilon_{3} $  &      -0.035         &       0.011         &   $$-$0.039^{*}$  &       0.001          \\
            &     (0.026)         &     (0.015)      &    (0.023)         &     (0.009)           \\
\addlinespace
$\mathrm{constant}$&      -$0.755^{**}$ &       0.076       &     -$0.789^{***}$&     -$0.231^{***}$ \\
            &     (0.350)         &     (0.110)           &(0.295)         &     (0.067)          \\
\midrule
\(N\)       &         910         &        2909           &  910         &        2909          \\
\(R^{2}\)   &       0.910         &       0.860           &  0.909         &       0.866           \\
\midrule
Country-\\
Industry FE   &        Yes         &        Yes         &        Yes         &        Yes         \\
Year FE   &        Yes         &        Yes         &        Yes         &        Yes         \\
\toprule
\end{tabular*}
\footnotetext{Note : This table shows the effects, which classifies importers on technological aspects: G5 and Non-G5. Group A is the worst seven of the 14 manufacturing industries in terms of the MAX value of centrality. Group B is the other 7. Standard errors in parentheses, \footnotesize{* \(p<0.1\), ** \(p<0.05\), *** \(p<0.01\)}}
\end{minipage}
\end{center}
\end{table}
The results are reported in Table 4. The results for non-G5 countries are similar to those for Europe. The effect from high centrality countries is the highest, and the effect from middle centrality countries is also significant. On the other hand, G5 countries are positively significantly affected by the R\&D contents from their home countries. It is confirmed that the effect from the domestic R\&D contents is not the nature of North America, but the nature of G5 countries with high technological capabilities. The results are generally consistent with the assumption, however the international spillover effects from high centrality countries are also observed in G5. Therefore, a country with high technological capability will benefit from the spillover effects if the trading country too has some extent of high level of technological. In addition, there is no common significant effects in Group A's industries between G5 and non-G5 countries.
\subsection{Discussion}
\hspace*{3mm}Based on the results (Table 3 and 4), I discuss the relationship between the spillover effects and centrality. In regards with the scale of the industries and exporters' centrality, the results in Table 3 show that there is a large difference in the international spillover effects between two group industries. First, in relatively low centrality industries (Group A), there are statistically insignificant effects from the foreign R\&D contents across all the classifications. Second, in more centralized industries (Group B), European countries have strong significance effects from exporters with high centrality because many countries participate in the EU, making it easy to trade within the EU, and the large number of hubs make trade more active. Further, I find that it is significant in countries with middle centrality as well. In Appendix, I further investigate the significance of Middle. I find that the demand from countries with middle centrality is on the rise, with a large increase in 2002 (Figure 3). In fact, the interaction term of the Middle centrality classification, with the dummy variable equal to 1, is positively significant after 2002 (Table 8). Specifically, the effects of the Middle categorization suggest that countries with middle centrality in imports and high centrality in exports are important (Figure 4), and there is a growing presence of mediators in trade. Finally, only North America shows different results in Table 3 that I further investigate. Instead of classifying the importers by geography, I now classify them by technology, i.e., G5 and non-G5. It turns out that G5 countries are positively significantly affected by R\&D contents from their home countries. The international spillover from countries with high centrality is also observed in G5 (Table 4).\\
\hspace*{3mm}In the next section, I conduct robustness checks for the results in Table 4, and explain the method and results of the robustness checks.
\subsection{Robustness checks}
\hspace{3mm}In this section, I conduct two robustness checks for the results reported in Table 3. First, I use the other classification of centrality to show that the results do not vary with the centrality classification used in this study. Second, I use one-year lagged explanatory variables for capturing the effects that are not reflected immediately after the trade.\\
\hspace*{3mm}In the analysis, since I use the classification of centrality categorized at a rate of 33\%, and examine the effects of each, I classify the centrality in the other classification, and investigate if the same results can be obtained. I define the top five countries as high centrality, the next top five countries as middle centrality, and the remaining countries as low centrality.
\begin{table}[h]
\begin{center}
\begin{minipage}{\textwidth}
\caption{Other centrality classification results}
\begin{tabular*}{\textwidth}{@{\extracolsep{\fill}}lcccccc@{\extracolsep{\fill}}}
\toprule%
& \multicolumn{3}{@{}c@{}}{Group A} & \multicolumn{3}{@{}c@{}}{Group B} \\\cmidrule{2-7}%
 &        (1)         &        (2)          &        (3)         &        (4)      &        (5)          &        (6)        \\
            &        All         &        Europe          &        North America         &        All         &        Europe          &        North America \\
\midrule
\rm{domestic}&      -0.023         &      -0.027         &      $0.052^{**}$        &      $-0.026^{*}$  &      -0.033         &       $0.033^{**}$ \\
            &     (0.014)         &     (0.018)         &     (0.024)         &     (0.014)         &     (0.020)         &     (0.015)         \\
\textbf{exporter's}\\
\textbf{centrality:}\\
\addlinespace
\hspace{3mm}\rm{High}   &       0.006         &      0.024         &       -0.055         &       $0.037^{**}$         &       $0.061^{**}$ &      -0.066         \\
            &     (0.012)         &     (0.021)         &     (0.047)         &     (0.016)         &     (0.020)         &     (0.058)         \\
\addlinespace
\hspace{3mm}\rm{Middle} &       0.041 &       0.070&       $0.042^{*}$         &       $0.028^{*}$ &       $0.030^{*}$         &      -0.031         \\
            &     (0.033)         &     (0.060)         &     (0.024)          &     (0.015)         &     (0.016)         &     (0.033)         \\
\addlinespace
\hspace{3mm}\rm{Low}    &       0.025         &       0.026         &      -0.001         &       0.016         &       0.008         &       0.072         \\
            &     (0.017)         &     (0.025)         &     (0.020)         &     (0.014)         &     (0.014)         &     (0.043)         \\
\addlinespace
$\mathrm{constant}$&       0.124         &       0.142         &       -0.168         &      -$0.133^{**}$ &               -$0.179^{**}$ &       0.258         \\
            &     (0.080)         &     (0.101)         &     (0.145)         &     (0.060)         &     (0.065)         &     (0.179)         \\
\midrule
\(N\)       &        3819         &        3091         &         546         &        3819         &        3091         &         546         \\
\(R^{2}\)   &       0.868         &       0.856         &       0.954          &       0.870         &       0.859         &       0.959         \\
\midrule
Country-\\
Industry FE   &        Yes         &        Yes         &        Yes         &        Yes         &        Yes         &        Yes                  \\
Year FE   &        Yes         &        Yes         &        Yes         &        Yes         &        Yes         &        Yes                 \\
\toprule
\end{tabular*}
\footnotetext{Standard errors in parentheses, \footnotesize{* \(p<0.1\), ** \(p<0.05\), *** \(p<0.01\)}}
\end{minipage}
\end{center}
\end{table}
The results are reported in Table 5, and show that the significance is almost the same. Only the significance of exporters with high forward centrality disappears in G5, but the other significances remain. Therefore, I find that changing the classification of centrality does not affect the results.\\
\hspace*{3mm}Next, the technology of the trading partners is not immediately reflected in their own industry, but may affect productivity across time. To account for this effect, a one-year lag is given to the explanatory variables in the analysis.
\begin{table}[h]
\begin{center}
\begin{minipage}{\textwidth}
\caption{1 year lagged results}
\begin{tabular*}{\textwidth}{@{\extracolsep{\fill}}lcccccc@{\extracolsep{\fill}}}
\toprule%
& \multicolumn{3}{@{}c@{}}{Group A} & \multicolumn{3}{@{}c@{}}{Group B} \\\cmidrule{2-7}%
 &        (1)         &        (2)          &        (3)         &        (4)      &        (5)          &        (6)        \\
            &        All         &        Europe          &        North America         &        All         &        Europe          &        North America \\
\midrule
\rm{domestic, lagged}&      -0.014         &      -0.017         &       $0.044^{**}$ &      -0.018         &      -0.020         &       $0.044^{**}$ \\
            &     (0.012)         &     (0.016)         &     (0.021)          &     (0.013)         &     (0.016)         &     (0.018)         \\
\textbf{exporter's}\\
\textbf{centrality:}\\
\addlinespace
\hspace{3mm}\rm{High, lagged}   &       $0.026^{*}$ &       0.012         &       0.011      &       $0.026^{*}$         &       $0.032^{*}$  &       0.022            \\
            &     (0.013)         &     (0.020)         &     (0.027)       &     (0.014)         &     (0.018)         &     (0.055)           \\
\addlinespace
\hspace{3mm}\rm{Middle, lagged} &       0.018         &       0.037         &      -0.031        &       0.025         &       $0.029^{*}$  &      -0.078  \\
            &     (0.017)         &     (0.025)         &     (0.021)        &     (0.014)         &     (0.016)         &     (0.063)          \\
\addlinespace
\hspace{3mm}\rm{Low, lagged}    &       0.006         &       0.001         &      -0.004        &       0.011         &       0.002         &       0.000          \\
            &     (0.010)         &     (0.012)         &     (0.013)        &     (0.008)         &     (0.011)         &     (0.022)          \\
\addlinespace
$\mathrm{constant}$&       0.053         &       0.057         &      -0.180        &      -0.082         &      -$0.137^{*}$  &       0.013          \\
            &     (0.077)         &     (0.097)         &     (0.160)        &     (0.063)         &     (0.076)         &     (0.231)           \\
\midrule
\(N\)       &        3525         &        2853         &         504        &        3525         &        2853         &         504          \\
\(R^{2}\)   &       0.883         &       0.872         &       0.963         &       0.885         &       0.875         &       0.964          \\
\midrule
Country-\\
Industry FE   &        Yes         &        Yes         &        Yes         &        Yes         &        Yes         &        Yes                  \\
Year FE   &        Yes         &        Yes         &        Yes         &        Yes         &        Yes         &        Yes                 \\
\toprule
\end{tabular*}
\footnotetext{Standard errors in parentheses, \footnotesize{* \(p<0.1\), ** \(p<0.05\), *** \(p<0.01\)}}
\end{minipage}
\end{center}
\end{table}
The results are reported in Table 6, and show that the significance is almost the same, and the effects are smaller. Therefore, I find that using the one-year lagged variable does not produce significantly different results from the hypothesis.
\section{Conclusion}\label{sec6}
\hspace*{3mm}In this paper, I examine the relationship between the international R\&D spillovers and the structure of the GVCs by using 21 countries and 14 manufacturing industries as samples for the period of 1995 to 2007. To this end, I identify the centrality of each industry in the GVCs in each year, using the world input-output table, and categorize the countries as per high, middle, and low centrality. Further, I classify the industries into two categories. Based on these, I examine the difference in the effects of exporters' centrality. I obtain the following results. First, I find that trading with high centrality countries have positively significant, and the highest effects, on domestic TFP. This result suggests that the high centrality countries have a greater role in transmitting knowledge, and diffusing effects. Second, I find that the exporters with middle centrality are also statistically significant. In particular, the increase in direct and indirect demand from countries with middle centrality since 2002 is believed to have contributed to the results. It is suggested that countries with middle centrality in imports and high centrality in exports are important for this result. Finally, it seems that only technologically-advanced countries (G5) are affected by the spillover effect from own. Looking at the share of domestic and foreign R\&D contents, the share of domestic content is much higher in the G5 countries than in the other countries. This result suggests that the proportion of the R\&D contents flowing into a country is important for R\&D spillovers.\\ 
\hspace*{3mm}This study is the first to consider, and empirically analyze, the position of the exporters in the GVCs for international R\&D spillovers. In contrast to the claims of Keller (1998), these findings suggest that the identity of both the importing and the exporting country is important in the international spillover effects. However, the impact of the final goods from domestic trade is significant only in the countries with a high share of domestic R\&D contents. In this study, I have only included Japan from the Asian countries due to data limitations, however, South-east Asia has built a strong production network, and I consider it possible to analyze the relationship between centrality and international spillover effects from a different perspective than that of Europe and North America. In addition to South-east Asia, there are many other countries, such as China and India, that have an important position in the GVCs, and if they can be included in the analysis, it will be possible to elucidate the relationship between the international R\&D spillover effects and the GVCs in more detail.
\section*{Appendix: Decomposition of the R\&D contents into direct and indirect demands}
\hspace{3mm}I decompose the foreign R\&D contents into direct and indirect, demands, and try to better understand the results in Table 4. Following Lumenga-Neso et al. (2005), I decompose it as follows:
\begin{align*}
    V^{f} =&\:\: V^{f\_\rm{direct}} + V^{f\_\rm{indirect}}\\
=&\:\: DB^{F}f \hspace{2.3mm}+ D[(I-B)^{-1}-(I+B)]^{F}f
\end{align*}
Table 7 displays the percentage of flows from countries in each centrality category for 1995 (panel (a)) and 2007 (panel (b)), respectively. As these panels indicate, most of them are from the High category (about 85\% and 73\% in panels (a) and (b), respectively). Although only 27\% of the countries are classified as High, we can see the importance of this category in the spillover effect. In addition, high indirect demand is common across categories. In ``Food, Beverages, and Tobacco,'' for example, there is a difference of about 50-60\%. Therefore, in the results of Table 4, we can see that indirect effects also account for a large part of the spillovers, and the results indicate that countries with High centrality are important as mediators of knowledge. On the other hand, comparing 1995 and 2007, it turns out that High has decreased and Middle has increased. To follow the trend in further detail, I analyze the time series change in the ratio of direct and indirect demands, respectively. As Figure 3 shows, the direct and indirect demands from countries with Middle centrality is on the rise, especially since 2001.\footnote{\;Although this figure shows the share in ``Electrical and Optical Equipment, '' the trend is basically the same for other industries as well.} This is one of the reasons why the effects from a country with Middle centrality are statistically significant.

\begin{table}[h]
\begin{center}
\begin{minipage}{\textwidth}
\caption{The shares of the R\&D contents}\label{Tab}
\begin{tabular*}{\textwidth}{@{\extracolsep{\fill}}lcccccc@{\extracolsep{\fill}}}
\toprule%
& \multicolumn{3}{@{}c@{}}{Direct ($V^{f\_\rm{direct}}$)(\%)} & \multicolumn{3}{@{}c@{}}{Indirect ($V^{f\_\rm{indirect}})$} \\\cmidrule{2-7}%
            classification\\
             of centrality:  & High   & Middle & Low & High     & Middle & Low \\
\midrule
(a) 1995:                                             &        &        &     &          &        &     \\
Food, Beverages, and Tobacco                & 12.7   & 2.3    & 0.1 & 71.1     & 13.3   & 0.5 \\
Textiles and Textile Products                & 30.2   & 3.9    & 0.3 & 54.7     & 10.6   & 0.3 \\
Leather, Leather, and Footwear                   & 24.3   & 3.8    & 0.3 & 59.5     & 11.8   & 0.4 \\
Wood and Products of Wood and Cork              & 21.5   & 3.7    & 0.2 &62.2     & 11.8   & 0.5 \\
Pulp, Paper, Printing, and Publishing & 25.6   & 4.3    & 0.2 & 58.5     & 11     & 0.4 \\
Coke, Refined Petroleum, and Nuclear Fuel        & 21.1   & 5.8    & 0.3 & 57.3     & 15.3   & 0.3 \\
Chemicals and Chemical Products                 & 48     & 6.9    & 0.5 & 36.9     & 7.3    & 0.3 \\
Rubber and Plastics                             & 45.5   & 6.1    & 0.5 & 39.8     & 7.8    & 0.3 \\
Other Non-Metallic Mineral                      & 28.4   & 5.2    & 0.3 & 55.2     & 10.6   & 0.4 \\
Basic Metals and Fabricated Metal            & 26.1   & 4.5    & 0.3 & 57.7     & 10.9   & 0.5 \\
Machinery, Nec                                  & 38.3   & 4.2    & 0.2 & 49.1     & 7.8    & 0.3 \\
Electrical and Optical Equipment             & 50.7   & 4      & 0.2 & 39.8     & 5.1    & 0.2 \\
Transport Equipment                          & 42.5   & 7.1    & 0.1 & 42.2     & 7.9    & 0.2 \\
Manufacturing, Nec; Recycling                & 27     & 4.1    & 0.3 & 57.2     & 10.9   & 0.4 \\
\\\vspace{2mm}
(b) 2007:                                             &        &        &     &          &        &     \\
Food, Beverages, and Tobacco                  & 11.9   & 4      & 0.2 & 60.4     & 22.5   & 1   \\
Textiles and Textile Products                & 27.9   & 9.2    & 0.4 & 44       & 17.9   & 0.6 \\
Leather, Leather, and Footwear                   & 22.9   & 7.5    & 0.6 & 49       & 19.3   & 0.7 \\
Wood and Products of Wood and Cork              & 19.3   & 7.8    & 0.5 & 51.8     & 19.8   & 0.8 \\
Pulp, Paper, Printing, and Publishing & 22.3   & 8.5    & 0.5 & 49.2     & 18.8   & 0.7 \\
Coke, Refined Petroleum, and Nuclear Fuel        & 16.1   & 6.3    & 0.5 & 52.9     & 23     & 1.1 \\
Chemicals and Chemical Products                 & 40.3   & 15.5   & 1   & 30.1     & 12.7   & 0.4 \\
Rubber and Plastics                             & 37.1   & 13.3   & 0.7 & 34.1     & 14.3   & 0.5 \\
Other Non-Metallic Mineral                      & 23.9   & 7.8    & 0.5 & 48.9     & 18.2   & 0.7 \\
Basic Metals and Fabricated Metal            & 21.8   & 6.3    & 0.4 & 52.3     & 18.4   & 0.8 \\
Machinery, Nec                                 & 35.2   & 7.9    & 0.5 & 40.7     & 15.2   & 0.5 \\
Electrical and Optical Equipment             & 39.7   & 12.4   & 0.4 & 32.9     & 14.4   & 0.3 \\
Transport Equipment                          & 40.2   & 6.5    & 0.3 & 39.5     & 13.1   & 0.4 \\
Manufacturing, Nec; Recycling                & 24.9   & 8      & 0.5 & 47.4     & 18.4   & 0.7\\
\toprule
\end{tabular*}
\footnotetext{Note: Author's calculation. This table shows the average percentage of foreign R\&D contents in the the final goods imported by each industry. The share is calculated as the average of the percentage of the total, including both direct and indirect demand.}
\end{minipage}
\end{center}
\end{table}
\begin{figure}[h]
      \begin{minipage}[b]{0.45\hsize}
        \centering
        \includegraphics[keepaspectratio, scale=0.5, clip]{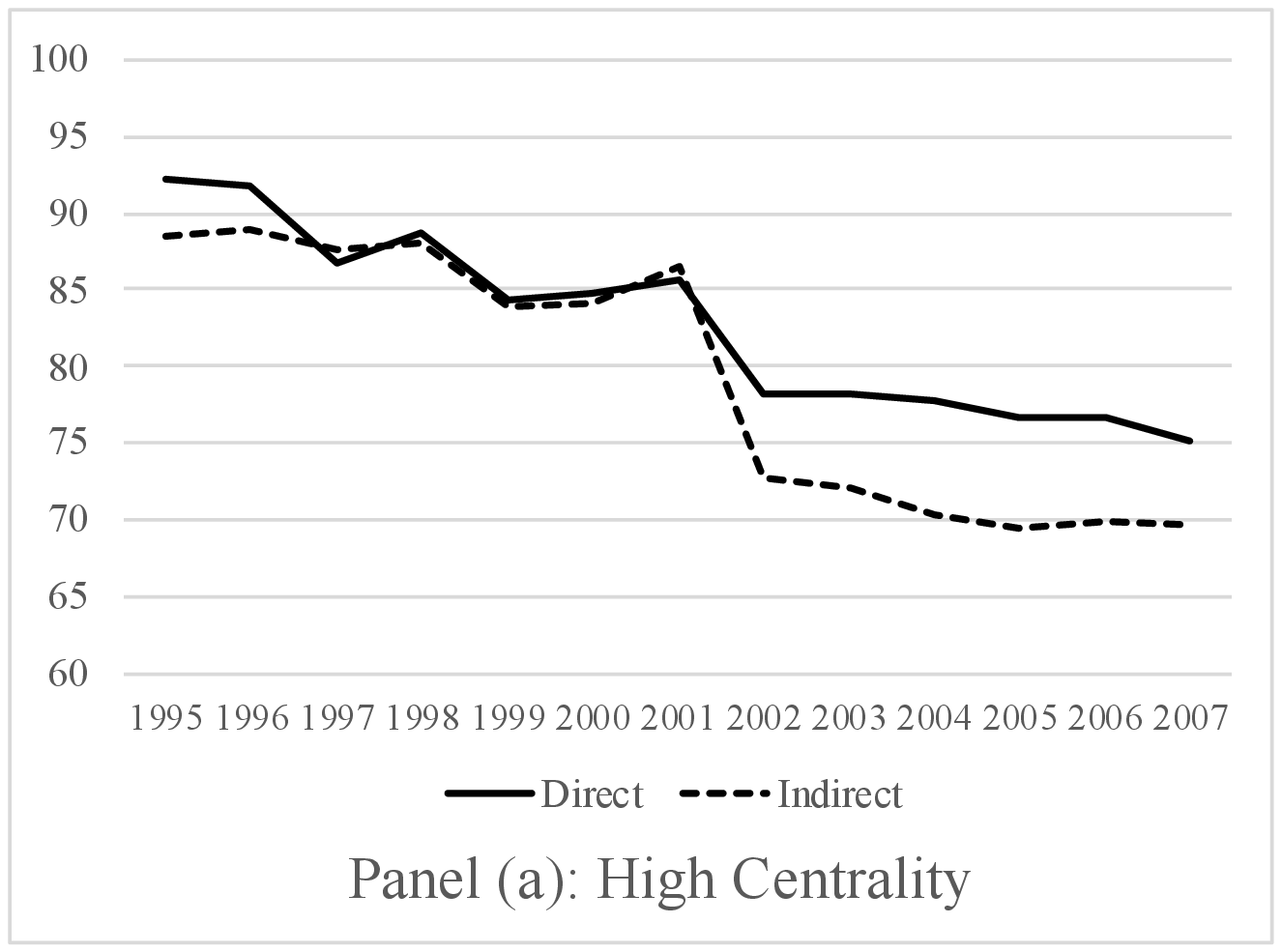}
      \end{minipage}
      \hspace{8mm}
      \begin{minipage}[b]{0.45\hsize}
        \centering
        \includegraphics[keepaspectratio, scale=0.5, clip]{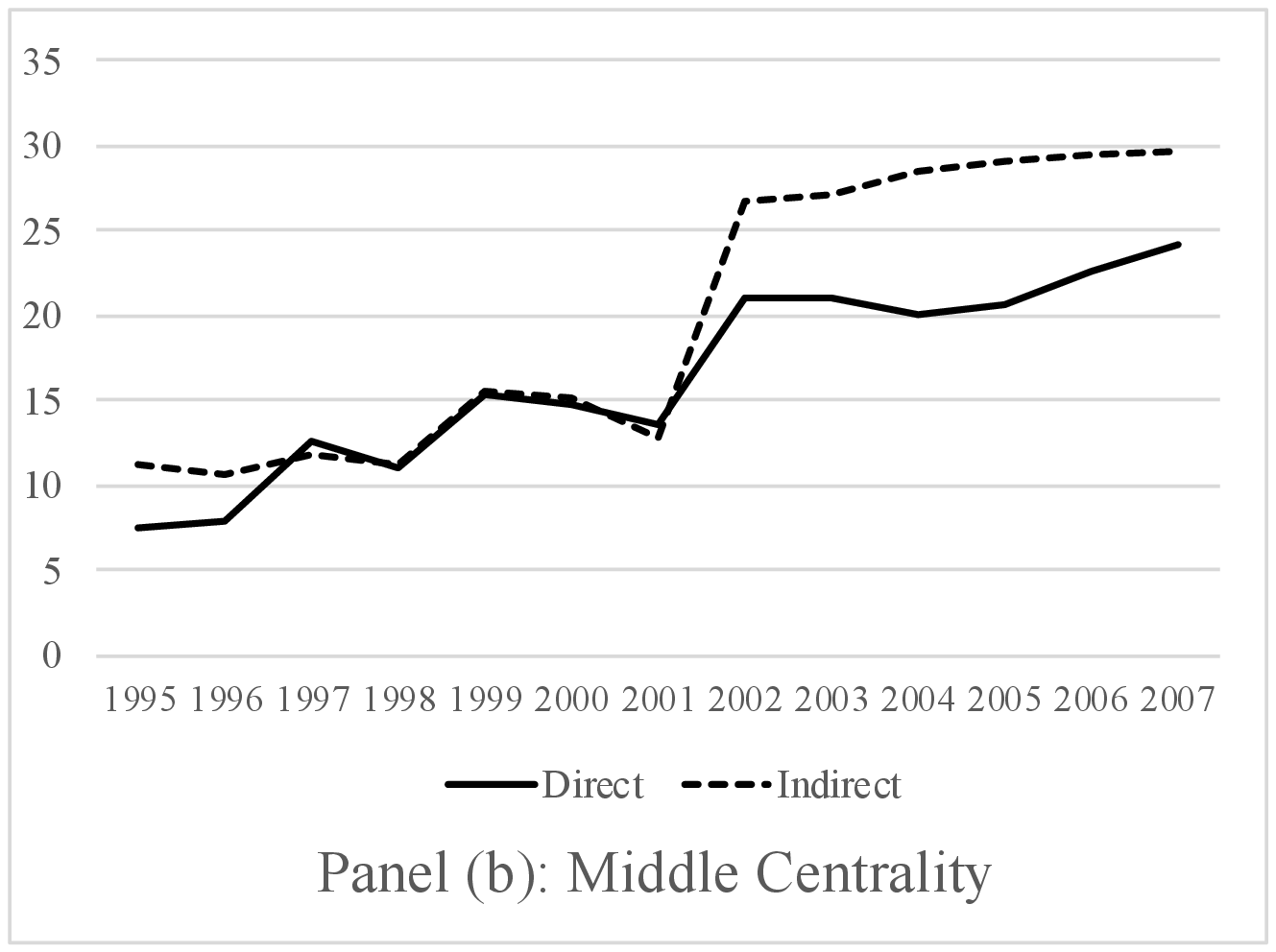}
      \end{minipage}
      \caption{The average R\&D contents share in Electrical and Optical Equipment}
\footnotesize{Source: Originally created by the author, using WIOD}\\
\footnotesize{Note: The y-axis represents the ratio of the foreign R\&D contents. The shares are calculated using direct and indirect demand, respectively, and as the average of all the sample countries.}
\end{figure}
Next, I add the dummy variable, that is equal to 1, for the years 2002-2007 ($Year^{02-07}$) as interaction term with centrality term, and estimate the change in the effect since 2002. Table 8 shows the results of the additional analysis, and rows 5-7 indicate the newly-added variables. The effects from the countries with Middle centrality are positive and significant post-2002. This result indicates that countries with Middle centrality are also important in the spillover effects since 2002. On the other hand, in all the samples, the effect from the countries with High centrality is negatively significant. These results correspond to the trend in Figure 3. In North America, there are no significant effects from the foreign R\&D contents, and it is affected only by the domestic R\&D contents.\\
\begin{table}[h]
\begin{center}
\begin{minipage}{\textwidth}
\caption{The results after adding the dummy variable}
\begin{tabular*}{\textwidth}{@{\extracolsep{\fill}}lcccccc@{\extracolsep{\fill}}}
\toprule%
& \multicolumn{3}{@{}c@{}}{Group A} & \multicolumn{3}{@{}c@{}}{Group B} \\\cmidrule{2-7}%
 &        (1)         &        (2)          &        (3)         &        (4)      &        (5)          &        (6)        \\
            &        ALL         &        EUR          &        NA         &        ALL         &        EUR          &        NA \\
\midrule
\rm{domestic}, $\beta$&      -0.022         &      -0.032         &       $0.040^{**}$ &      -$0.030^{*}$  &      -$0.042^{*}$  &       $0.046^{**}$ \\
            &     (0.015)         &     (0.020)         &     (0.020)         &     (0.017)         &     (0.020)         &     (0.021)         \\
\textbf{exporter's}\\
\textbf{centrality:}\\
\addlinespace
\hspace{3mm}\rm{High}, $\rho_{1}$   &      $0.037^{**}$ &       0.029         &      -0.030        &       $0.056^{**}$ &       $0.059^{**}$ &      -0.002          \\
            &     (0.015)         &     (0.024)         &     (0.024)        &     (0.021)         &     (0.023)         &     (0.048)          \\
\addlinespace
\hspace{3mm}\rm{Middle}, $\rho_{2}$ &       0.022         &       0.020         &      -0.017        &       0.015         &       $0.028^{*}$  &      -0.046          \\
            &     (0.019)         &     (0.027)         &     (0.025)         &     (0.013)         &     (0.015)         &     (0.069)         \\
\addlinespace
\hspace{3mm}\rm{Low}, $\rho_{3}$    &       0.014         &       0.010         &       0.008         &       $0.018^{*}$  &       0.013         &       0.001         \\
            &     (0.012)         &     (0.015)         &     (0.024)         &     (0.009)         &     (0.011)         &     (0.044)          \\
\addlinespace
\hspace{3mm}\rm{High\:*\:$Year^{02-07}$}&      -0.016         &      -0.019         &       0.048 &      -$0.024^{**}$ &       0.021         &       0.039        \\
            &     (0.010)         &     (0.019)         &     (0.041)       &     (0.011)         &     (0.028)         &     (0.111)           \\
\addlinespace
\hspace{3mm}\rm{Middle\:*\:$Year^{02-07}$}&       $0.026^{*} $ &       0.028         &      0.043 &       $0.040^{**}$ &       $0.045^{**}$          &      -0.050         \\
            &     (0.012)         &     (0.021)         &     (0.037)        &     (0.015)         &     (0.023)         &     (0.130)          \\
\addlinespace
\hspace{3mm}\rm{Low\:*\:$Year^{02-07}$}&      -0.006         &      -0.007         &      -0.044         &      -0.015         &      -$0.031^{*}$  &      -0.004         \\
            &     (0.007)         &     (0.008)         &     (0.033)        &     (0.009)         &     (0.015)         &     (0.039)          \\
\addlinespace
$\mathrm{constant}$&       0.105         &       0.133         &      -0.140        &      -$0.144^{**}$ &      -$0.192^{**}$ &       0.018          \\
            &     (0.094)         &     (0.120)         &     (0.140)        &     (0.064)         &     (0.081)         &     (0.214)          \\
\midrule
\(N\)       &        3819         &        3091         &         546        &        3819         &        3091         &         546  \\
\(R^{2}\)   &       0.868         &       0.856         &       0.960       &       0.872         &       0.861         &       0.958           \\
\midrule
Country-\\
Industry FE   &        Yes         &        Yes         &        Yes         &        Yes         &        Yes         &        Yes                  \\
Year FE   &        Yes         &        Yes         &        Yes         &        Yes         &        Yes         &        Yes                 \\
\toprule
\end{tabular*}
\footnotetext{Standard errors in parentheses, \footnotesize{* \(p<0.1\), ** \(p<0.05\), *** \(p<0.01\)}}
\end{minipage}
\end{center}
\end{table}
\hspace*{3mm}Furthermore, I identify the Middle categories that impact the results. In Table 7, I checked the respective percentages of the R\&D contents in High, Middle, and Low, however, here I will focus on Middle alone. I do not distinguish between direct and indirect demand, and consider the total demand. In ``Electrical and Optical Equipment,'' the time-series change in each category of Middle, averaged over the sample countries, as shown in Figure 4.\\
\begin{figure}[h]
  \begin{center}
     \includegraphics[width=0.73\textwidth]{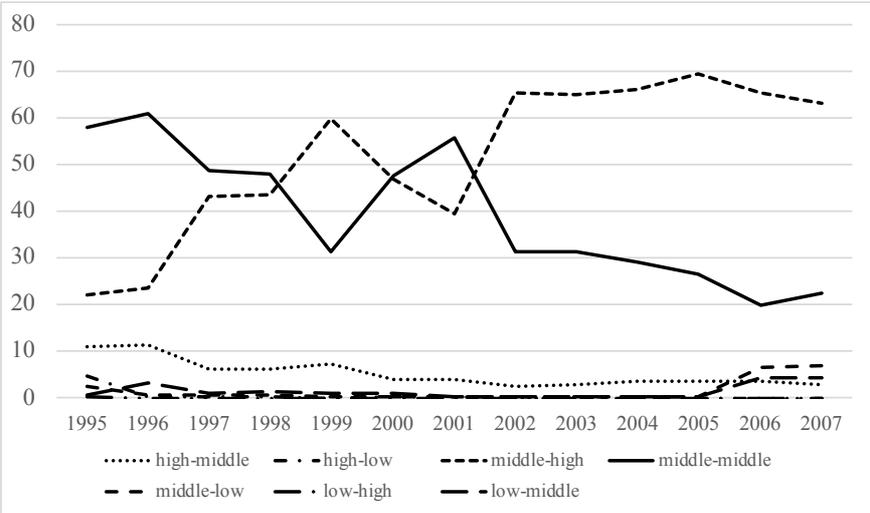}
   \end{center}
\caption{The average R\&D contents shares in Electrical and Optical Equipment in Middle classification}
\footnotesize{Source: Originally created by the author, using WIOD}\\
\footnotesize{Note: The y-axis represents the ratio of the foreign R\&D contents. The shares are calculated using the total demand, and the average of all the sample countries. For each category, for example, ``high-middle'' means that countries with high centrality in backward and middle centrality in forward.}
\end{figure}
As Figure 4 indicates, ``middle-high'' and ``middle-middle'' account for most of the Middle classification (about 80\%-97\%). In addition, while ``middle-middle'' is on a downward trend, ``middle-high'' is on an upward trend.\footnote{\;Although this figure shows the share in ``Electrical and Optical Equipment,'' the trend is the same for the other industries.} Therefore, when combined with the results in Table 8, we see that countries with middle centrality in imports but high centrality in exports also play an important role in the foreign R\&D spillovers as mediators. They are also important in the diffusion of knowledge because of their high centrality in exports.
\clearpage

\end{document}